\documentclass{PoS}
\usepackage{amsfonts}
\usepackage{amssymb}
\usepackage{textcomp}
\everymath{\displaystyle}
\usepackage{subfigure}

\title{Phase structures in fuzzy geometries}
\ShortTitle{Phase structures in fuzzy geometries}
\author{\speaker{T R Govindarajan}\\
Institute of Mathematical Sciences,\\
 Chennai, 600113, India\\
\email{trg@imsc.res.in}}

\author{S. Digal,\\
Institute of Mathematical Sciences,\\
 Chennai, 600113, India \\
\email{digal@imsc.res.in}}
 
\author{Kumar S. Gupta,\\
Theory Division, Saha Institute of Nuclear Physics,\\
1/AF Bidhannagar, Calcutta 700064, India \\
\email{kumars.gupta@saha.ac.in}}

\author{X. Martin \\
LMPT, UFR Sciences et Techniques, Universite de Tours,\\
Parc de Grandmont, 37200 TOURS, France \\
\email{xavier@lmpt.univ-tours.fr}}

\FullConference{Proceedings of the Corfu Summer Institute 2011 School and 
Workshops on Elementary Particle Physics and Gravity\\
September 4-18, 2011\\
Corfu, Greece}

\abstract{We study phase structures of quantum field theories in fuzzy geometries.
Several examples of fuzzy geometries as well as QFT's on such geometries are considered. 
They are fuzzy spheres and beyond as well as noncommutative deformations of BTZ blackholes.
Analysis is done  analytically and through simulations. Several features like 
novel stripe phases as well as spontaneous symmetry breaking 
avoiding Colemen, Mermin, Wagner theorem are brought out. Also we establish that 
these phases are stable due to topological obstructions.}

\begin{document}

\def\be{\begin{equation}}
\def\ee{\end{equation}}
\def\beq{\begin{eqnarray}}
\def\eeq{\end{eqnarray}}
\def\bn{\begin{eqnarray*}}
\def\en{\end{eqnarray*}}
\def\slas{\!\!\!/}
\def\BI{{\rm 1\!l}}
\def\P{\Phi}
\def\p{\phi}
\def\w{\omega}
\def\W{\Omega}
\def\O{{\cal{O}}}
\def\a{\alpha}
\def\b{\beta}
\def\s{\sigma}
\def\S{\Sigma}
\def\d{\delta}
\def\D{\Delta}
\def\g{\gamma}
\def\t{\theta}
\def\T{\Theta}
\def\G{\Gamma}
\def\z{\zeta}
\def\Z{\Psi}
\def\pd{\partial}
\def\e{\epsilon}
\def\n{\eta}
\def\m{\mu}
\def\r{\rho}
\def\t{\theta}
\def\R{\Rho}
\def\bra{{\langle}}
\def\ket{{\rangle}}
\def\bp{{\bf p}}
\def\bq{{\bf q}}
\def\bk{\bf k}
\def\br{{\bf r}}
\def\bx{{\bf x}}
\def\by{{\bf y}}
\def\l{\lambda}
\def\L{\Lambda}
\def\cL{{\cal{L}}}
\def\cH{{\cal{H}}}
\def\cP{{\cal{P}}}
\def\cU{{\cal{U}}}
\def\cN{{\cal{N}}}
\def\cT{{\cal{T}}}
\def\cC{{\cal{C}}}
\def\cD{{\cal{D}}}
\def\cJ{{\cal{J}}}
\def\cK{{\cal{K}}}
\def\cM{{\cal{M}}}
\def\cA{{\cal{A}}}
\def\cB{{\cal{B}}}
\def\cO{{\cal{O}}}
\def\cR{{\cal{R}}}
\def\cG{{\cal{G}}}
\def\cS{{\cal{S}}}
\def\cF{{\cal{F}}}
\def\cI{{\cal{I}}}
\def\cZ{{\cal{Z}}}
\def\la{\langle}
\def\ra{\rangle}

\newcommand{\diff}{\mathrm{d}}

\section{Introduction}
General relativity and quantum mechanics together imply that space-time structure at the Planck scale is not described 
by conventional notions of geometry. This was pointed out by many. In particular see Doplicher etal., \cite{dop2} as well as 
Werner Nahm (see \cite{podles}). These come about due to appearance of null horizons. Interestingly extra dimensional spacetimes
with higher dimensional `Planck scale'  $O(Tev)$ will require the extra dimensional space description to be fuzzy.
Surprisingly the difficulties in defining geometry at infinitesimal distances
were anticipated much earliar: {\it{`it seems that empirical notions on which the metrical determinations
of space are founded, the notion of a solid body and a ray of light cease
to be valid for the infinitely small. We are therefore quite at liberty
to suppose that the metric relations of space in the infinitely small do
not conform to hypotheses of geometry; and we ought in fact to suppose it,
if we can thereby obtain a simpler explanation of phenomena': Riemann}}, \cite{clifford}

Field theories on non-commutative geometries are inherently non-local leading to mixing of the infrared and
ultraviolet scales. This, in turn, is responsible for new ground
states with spatially varying condensates. Many non-perturbative
studies have established that non-commutative spaces, such as the
Groenewold-Moyal plane and fuzzy spheres, allow for the formation of
stable non-uniform condensates as ground states.  Exploring such 
implications of the non-local nature of field theories is very important
in many areas of quantum physics. \cite{pinzul,gubser}
 
Since different phases are intimately connected with spontaneous
symmetry breaking (SSB), the role of symmetries in noncommutative
geometries themselves is subtle.  This issue is important in 2D
because the Coleman-Mermin-Wagner (CMW) theorem states that there can
be no SSB of continuous symmetry on 2-dimensional commutative
spaces. There is no obvious generalisation of the CMW theorem for
non-commutative spaces, since the theorem relies strongly on the
locality of interactions.  Non-commutative spaces admit non-uniform
solutions (in the mean field) and one can ask the question what
happens to the stability of these configurations. Non-uniform
condensates naturally have an infra-red cut-off for the
fluctuations. This cut-off softens the otherwise divergent
contributions of the Goldstone modes\cite{xavier,denjoe,digal1,digal2,
bietenholz,ambjorn,panero,medina}.

There have been various attempts to study gravity theories within the noncommutative framework \cite{wess1,seckin}. 
This has led to a Hopf algebraic description of noncommutative black holes \cite{brian1,schupp} and 
FRW cosmologies \cite{ohl}. A large class of such black hole solutions, including the noncommutative BTZ \cite{btz1,btz2} 
and Kerr black holes, exhibits an universal feature where the Hopf algebra is described by a noncommutative 
cylinder \cite{petergrosse}, which belongs to the general class of the $\kappa$-Minkowski algebras \cite{L1,L4,M1,M3}. 
we shall take the noncommutative cylinder and the associated algebra as a model for noncommutative black holes. 

The study of quantum field theories in the background of black holes has led to the discovery of 
interesting features associated with the underlying geometry, such as the Hawking radiation and black hole entropy. 
In the noncommutative case, the black hole geometry is replaced with the algebra defined by the noncommutative cylinder. 
In order to probe the features of a noncommutative black hole, it is useful to analyze the behaviour of a quantum fields 
coupled to the noncommutative cylinder algebra. Scalar field theories have been extensively studied on 
$\kappa$-Minkowski spaces \cite{luk2,luk3,G1,G2,ksms}, which has led to twisted statistics and deformed oscillator 
algebra for the quantum field \cite{G1,G2,dice2010}. Theories on the noncommutative cylinder lead to 
quantization of the time operator \cite{C2,paulo}. See Madore \cite{madore} or Balachandran et al \cite{balbook}
for an introduction to fuzzy geometries.

In this paper we discuss examples of fuzzy geometries in Sec 2. In Sec 3 
we consider QFT's on such geometries. Following this in Sec 4 we take up 
aspects of numerical simulations of flutuations of fields on fuzzy geometries
and present our results. Lastly we conclude with discussions 
on implications of our results in Sec 5.

\section{Examples of fuzzy geometry}
It has been well known that representations of $SU(2)$ Lie-algebra provide 
a basis for the study of functions on a fuzzy spheres. This can be understood 
by the quantisation of coadjoint orbits ${SU(2)}/{U(1)}$. The generators 
of the Lie algebra $X_i$ satisfy: 
\be 
[X_i,X_j]~=~i\e_{ijk}X_k, \qquad \sum X_i^2 ~=~R^2.
\ee
The fact $S^2~=~CP^1$ is a coadjoint orbit is useful in quantising
this space. This can be extended to $CP^2~=~{SU(3)}/{SU(2)\otimes U(1)}$
and any $CP^n$.  
The fuzzy torus is defined by the two generators $U,V$ satisfying $U~V~=~e^{i\t}~V~U$. 
Finite dimensional representations
can easily be constructed for this algebra for rational $\t$.
We will explain certain non standard ones.
\subsection{Higgs algebra}
The Higgs algebra is defined by $[X_+,X_-]~=~\a Z + \b Z^3, ~[X_\pm,Z]~=~\mp X_+$.
It can be easily checked that the Casimir for this algebra is given by:
\be
\mathcal{C} = \frac{1}{2}\left[\{X_+, X_-\} + g(Z)
+ g(Z-1)\right],
\ee
where $g(Z)$ is
\be
g(Z) = C_0 + \frac{\a}{2} \, Z (Z+1) + \frac{\b}{4} \,
Z^2 (Z+1)^2 .
\ee
For $C_0 = \mu^2$, $\a=-2(2\mu~+~1)$, and $\b=4$ The Casimir reduces to the expression:
$X^2 + Y^2 + (Z^2 ~-~ \mu)^2$.
Equating the Casimir to 1 and plotting the function for different values of $\mu$
we see interesting topology change \cite{govind}. 
\begin{figure}[hbt!]
\begin{center}
    { \label{f:subfig-1}
      \includegraphics[height=1.5in,width=1.5in]{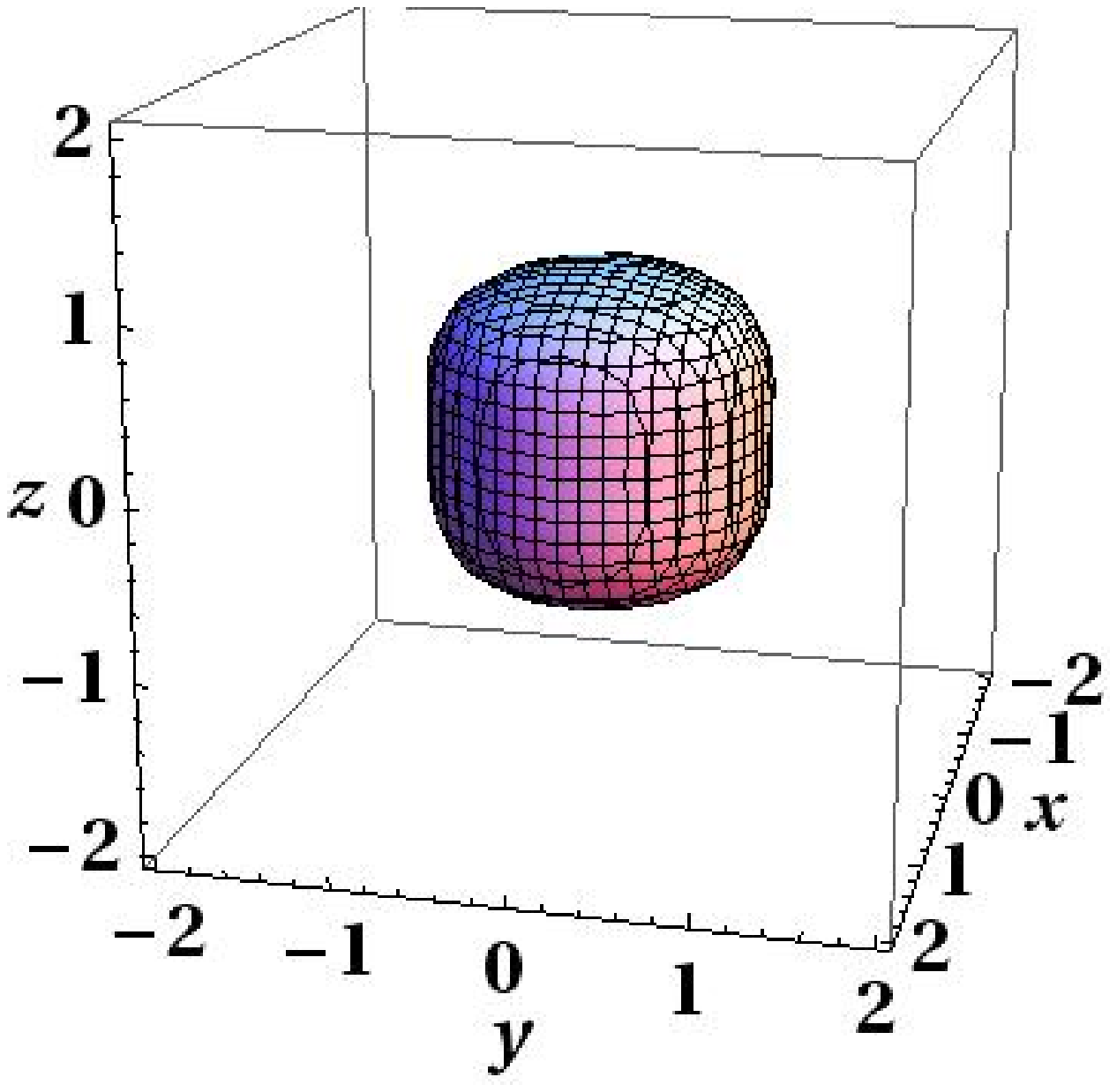} }
    { \label{g:subfig-2}
      \includegraphics[height=1.5in,width=1.5in]{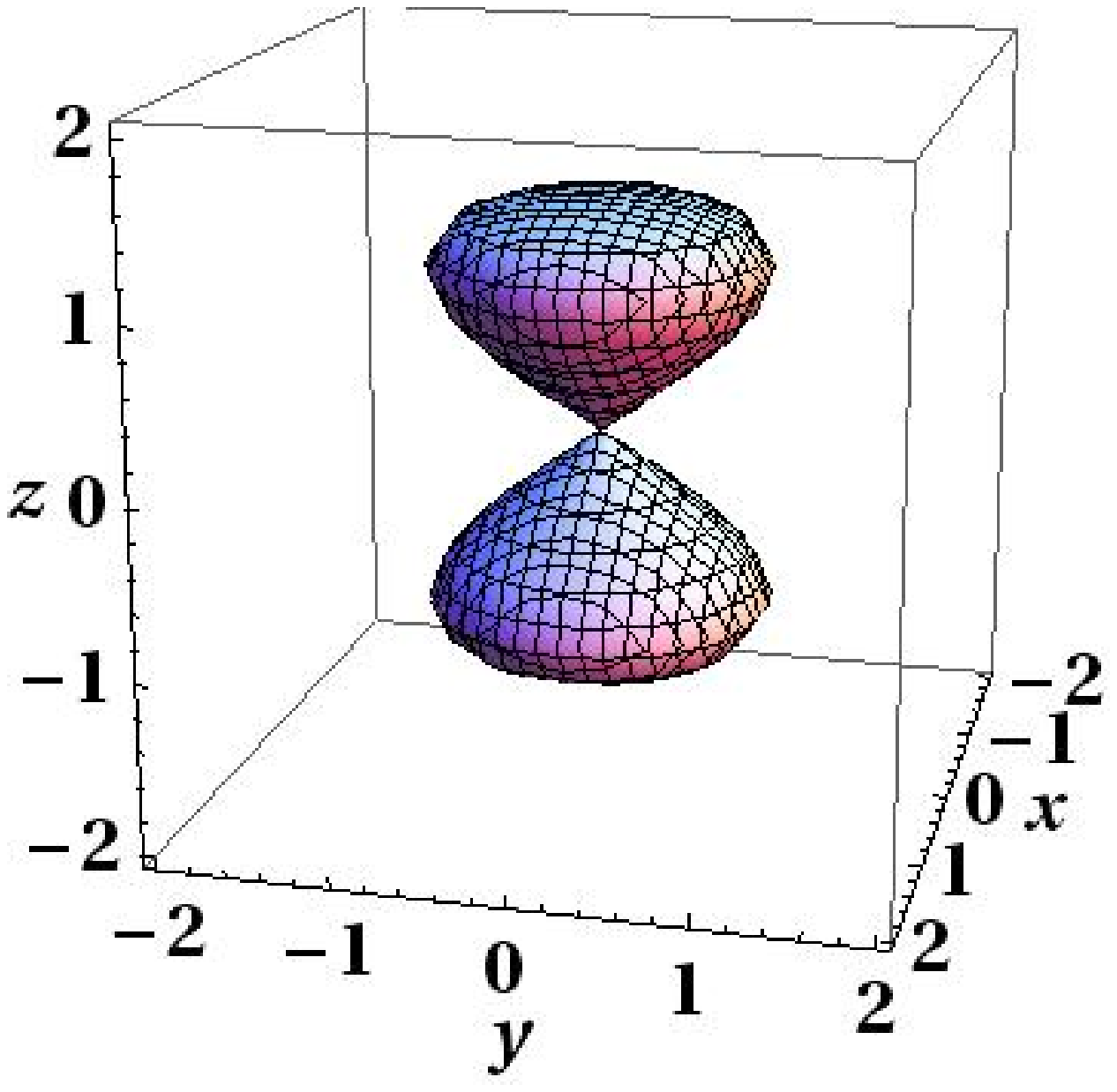} }
    { \label{h:subfig-2}
      \includegraphics[height=1.5in,width=1.5in]{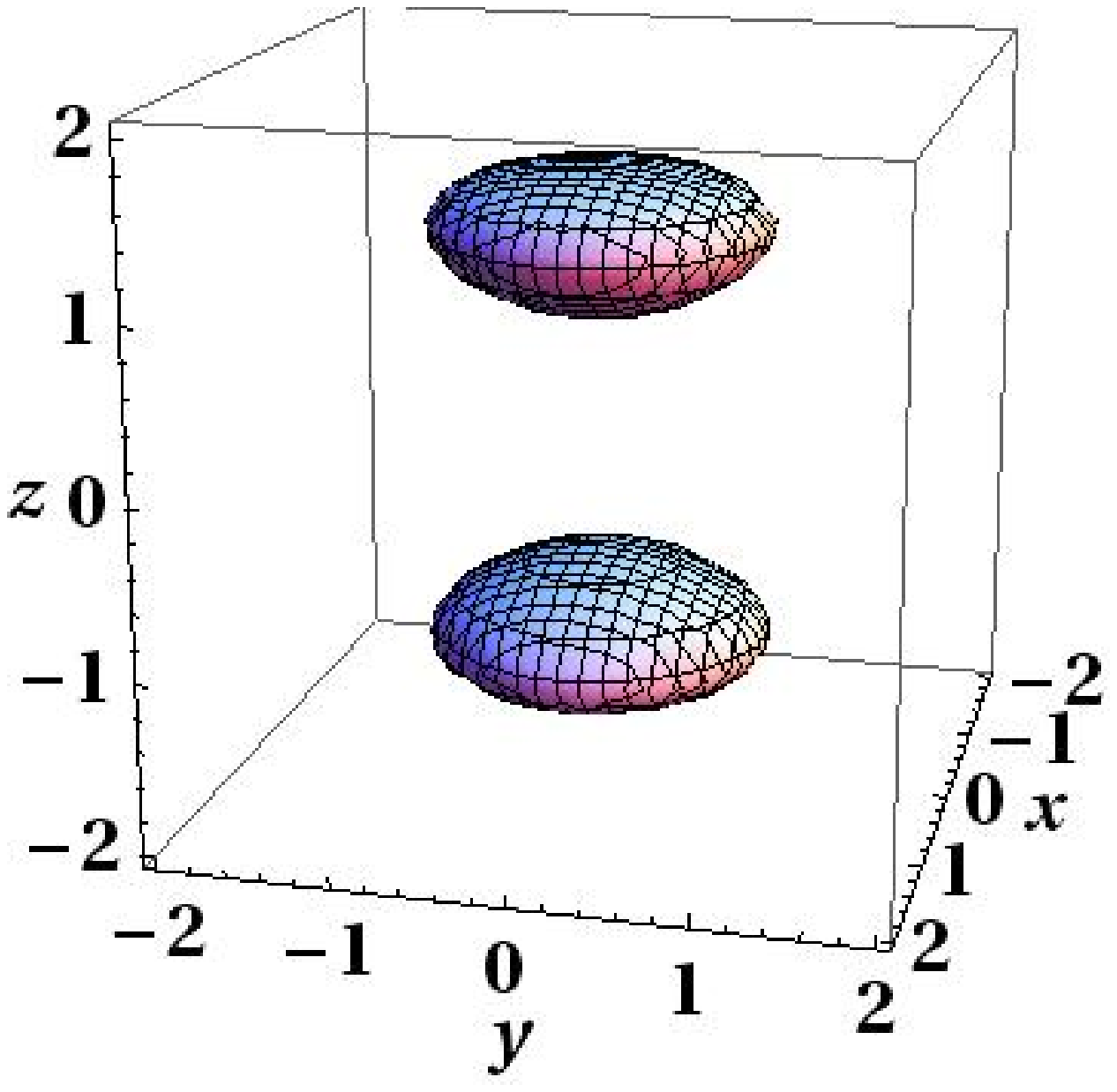} }
\end{center}
\caption{Surface plots depicting the change in topology for {$\mu = 0,1,1/2$}.}
\end{figure}
Such changes in the topology were first considered  by Arnlind  etal \cite{hoppe}.

\subsection{Fuzzy cylinder}
The NC cylinder is defined by the relation
\be
\left[~Z,~e^{i \phi}~\right]~=~\a e^{i \phi}
\label{angle}
\ee
where $Z$ is hermitian and $e^{i \phi}$ is unitary.   
Since we are interested in simulations, we have to discretise the above NC cylinder.

In the rest of this paper, we will work with $\alpha=1$ without loss
of generality, since the simple scaling $Z\rightarrow Z/\alpha$ can
scale $\alpha$ away in the commutation relation (\ref{angle}).

For this purpose consider the spin $J$ irreducible representation (IRR) 
of the $SU(2)$ Lie algebra, given by
\be
[X_+,X_-]~=~2X_3, ~~[X_\pm, X_3]~=~\mp X_\pm. \label{CR}
\ee
Since the operator $Z$ generates rotations around the axis of the cylinder, it can be identified with $X_3$.
But when we use the finite dimensional representations
of $SU(2)$ we cannot implement (\ref{angle}) with unitarity for $e^{i \phi}$.
For this purpose, we decompose $X_+$ as product of a unitary and a Hermitian
operator as given by
\be
X_+~=~e^{i\phi}~R \label{defRe}
\ee
Here $R$ is necessarily singular and  can not be inverted. However a partial inverse
$\tilde{R}$ can be found such that $R \tilde{R}~=~P$, the projector such that
$1-P$ projects to the kernel of $R$. Thus we get
\be
~[Z, e^{i \phi}]~P=~ e^{i \phi}~P\label{Pcr}
\ee

To find a representation for $R$ and $e^{i \phi}$, we can look at
\beq
R^2=X_-X_+=\vec{L}^2-L_3(L_3+1)
\eeq
which commutes with $Z=L_3$. Remember that in the usual representation of angular momentum,
$L_3|l,m>=m|l,m>$ with $|m|\leq l$. Shifting the indices from $0$ to
$2l+1=2J$, we have $j=m+l+1$, leading to
\begin{eqnarray*}
X_-X_+|l,m> & = & [l(l+1)-m(m+1)]|l,m>=[(l+1/2)^2-(m+1/2)^2]|l,m>\\
X_-X_+|J,j> & = & [J^2-(J-j)^2]|J,j>=j(2J-j)|J,j> \end{eqnarray*}
There is only one hermitian positive solution to this equation which takes the
form
\be
R_{ij}~=~\sqrt{i(2J-i)}~\d_{i,j}
\ee
which is diagonal as expected, and whose null space is along the top
state $|J,2J>$. As a result, $P=1-|J,2J><J,2J|$, and
$$\tilde{R}=\sum_{i=1}^{2J-1} [i(2J-i)]^{-1/2}\,|J,i><J,i|$$
where the sum stops at $i=2J-1$ so that there is a zero in the last position on
the diagonal.
It is now possible to deduce the first $2J-1$ lines of the unitary matrix
$e^{i \phi}$ from (\ref{defRe}):
$$\left.\begin{array}{rcl} X_+|J,j> & = & \sqrt{j(2J-j)}|J,j+1>\\
X_+|J,j> & = & e^{i\phi}~R|J,j>=\sqrt{j(2J-j)}e^{i\phi}~|J,j>
\end{array}\right\} \Rightarrow e^{i\phi}~|J,j>=|J,j+1>,\ j<2J.$$
Eq. (\ref{defRe}) yields no equation for the last column which is
instead determined from its unitarity. The columns
$1,\cdots,2J-1$ of $e^{i \phi}$, given by $|J,j+1>$, form an orthonormal set,
as expected for a unitary matrix. Then the last
column will be a vector orthogonal to all these vectors and thus can only be
proportional to $|J,1>$. After normalisation that still leaves a $U(1)$
freedom so that
\be
(e^{i \phi})_{ij}~=~ \d_{i,j+1}~+~ e^{i\b}~\d_{i,1}\d_{j,2J}
\label{newangle}
\ee
where $\b$ can be any real number. For $\b=0$, $e^{i \phi}$ is just a
circular permutation of length $2J$.

\subsection{Noncommutative BTZ blackhole}

We briefly summarize the essential features of a noncommutative black hole which 
is useful for our analysis. In the commutative case, 
a non-extremal BTZ black hole is described in terms of the coordinates 
$(r, \phi, t)$ and is given by the metric \cite{btz1,btz2}
\be 
ds^2= \biggl( M - \frac {r^2}{\ell^2} - \frac{J^2}{4r^2}\biggr)
dt^2  - \biggl( M - \frac {r^2}{\ell^2} -
\frac{J^2}{4r^2}\biggr)^{-1} dr^2+ r^2 \biggl(d\phi - \frac J{2r^2} dt\biggr)^2
\;,
\label{btzmtrc}
\ee
where $0\le r<\infty\;, \;-\infty < t<\infty\;, \;0\le \phi<2\pi\;,$ $M$ and
 $J$ are respectively the mass and spin of the black hole, and  $\Lambda= -1/\ell^2$ is the 
cosmological constant. In the non-extremal case, the two distinct horizons $ r_{\pm}$ are given by
\be 
r_\pm^2 =\frac {M\ell^2}2 \biggl\{ 1\pm \bigg[ 1 - \biggl(\frac
J{M\ell}\biggr)^2\biggr]^{\frac 12} \biggr\}. \;
\label{rpm}
\ee

An alternative way to obtain the geometry of the BTZ black hole is to
quotient the manifold $AdS_3$ or $SL(2,R)$ by a discrete subgroup of
its isometry. The noncommutative BTZ black hole is then obtained by a
deformation of $AdS_3$ or $SL(2,R)$ which respects the quotienting
\cite{brian1}.  In the noncommutative theory, the coordinates $r$,
$\phi$ and $t$ are replaced by the corresponding operators $\hat r$,
$\hat \phi$ and $ Z $ respectively, that satisfy the algebra
\be 
[\hat{t}, e^{i\hat \phi}] = \alpha e^{i\hat \phi}\qquad [\hat r,\hat t] =  
[\hat r,e^{i\hat \phi}] =0\;,
\label{qntmalg} 
\ee
where the constant $\alpha$ is proportional to $\ell^3/(r_+^2-r_-^2)$. We shall henceforth refer to (\ref{qntmalg}) 
as the noncommutative cylinder algebra. Furthermore, $\hat t$  denoting the operator corresponding 
the the axis of the cylinder, it will be therefore identified as the operator $Z$ in the following sections.

It may be noted that the operator $\hat r$ is in the center of the algebra (\ref{qntmalg}). In addition, 
it can be shown easily that $e^{-2\pi i\hat t/\alpha}$ belongs to the center of (\ref{qntmalg}) as well. 
Hence, in any irreducible representation of (\ref{qntmalg}), the element $e^{-2\pi i\hat t/\alpha}$ is proportional 
to the identity,
\be
e^{-2\pi i\hat t/\alpha} = e^{i\gamma}\BI,
\label{central}
\ee
where $\gamma \in R$ mod $(2 \pi)$. Eqn. (\ref{central}) implies that in any irreducible 
representation of (\ref{qntmalg}), the spectrum of the time operator $\hat{t}$, or $Z$, 
is quantized \cite{petergrosse,C2,paulo} and is given by
\be 
{\rm {spec}}~ {\hat t} = n \alpha
-\frac {\gamma\alpha}{2\pi}\;,\;\;n\in
{\mathbb{Z}}.
\label{dsctsptmt}
\ee
In what follows we shall set $\gamma = 0$ without loss of generality.

The noncommutative cylinder algebra (\ref{qntmalg}) belongs to a special class of $\kappa$ - Minkowski 
algebra and it appears in the description of noncommutative Kerr black holes \cite{schupp} and FRW cosmologies \cite{ohl}. 
We shall henceforth consider (\ref{qntmalg}) as a prototype of the noncommutative black hole.

\section{Scalar fields on fuzzy spheres/cylinder}
We shall now present our analytic and numerical analysis of scalar fields
on different fuzzy geometries.

Let $\Phi$ be a scalar field on a fuzzy sphere defined by spin $j~=~{(N-1)}/{2}$
representation. It is given by a $N \times N$ matrix. We consider the action given by:
\be 
S~=~\frac{4\pi}{N}~Tr~\{\Phi[L_i[L_i,\Phi]\}~+~R^2\{r\Phi^2~+~\l \Phi^4 ~\}
\ee
It is easy to see the ground states are characterised by 
\be 
\Phi = 0,~~ and \qquad \Phi \neq 0, ~~but~~  Tr~\Phi~=~0
\ee
which corresepond to the uniform and nonuniform or stripe phases.
We can obtain the continuum limit by taking $N \longrightarrow \infty$. 
The planar limit is obtained by taking  $R \longrightarrow \infty$.
One gets commutative planar or noncommutative planar (Moyal) limit depending
on ${R^2}/{N} \longrightarrow \infty$ or finite.
If we have a complex scalar field $\Phi$ then global $U(1)$ symmetry
can be broken contrary to the expectation from Coleman-Mermin-Wagner theorem in the NC limit.
This is due to the nonlocality of NC geometries. This has been shown 
through simulations in \cite{digal2}.

\subsection{Topological aspects on fuzzy spheres}
But our interest in this work is to consider the topological aspects of nonlinear 
fields  on fuzzy geometries. For this, we consider three hermitian scalar fields $\Phi_i$ with global $O(3)$ 
symmetry. The most general action upto quartic interactions takes the form,
\begin{equation}
S(\Phi) = \frac{4\pi}{N}\,{\rm Tr}\left[
\sum_i |\left[L_i,\Phi\right]|^2
+R^2\left(r|\Phi|^2+ i\beta \epsilon_{ijk}\Phi_i\Phi_j\Phi_k+\lambda(|\Phi|^2)^2+
\mu|\left[\Phi_i,\Phi_j\right]|^2\right)\right]
\label{action}
\end{equation}
In the mean-field the above theory admits a uniform condensate for $r<0$.
However the fluctuations of the Goldstone mode render
this solution unstable. Apart from the uniform condensate the above model
admits many meta-stable solutions.
To simplify our arguments we consider the case $\beta~=~0, \mu~=~0$:
We are interested on those solutions 
which are stable due to topological obstructions. For example,
\begin{equation}
\Phi_i = ~\alpha~ L_i,~~ \rm{with} ~~\alpha = \sqrt{\frac{\frac{2|r|}{\lambda}}{N^2-1}}
\label{winding}
\end{equation}
The analog of this configuration in continuum space is the hedgehog
configuration where the $O(3)$ spin vector on the sphere is pointing
radially outward. The spin is parallel to the position vector on
the sphere. This configuration is topologically stable as
it cannot be smoothly deformed to a uniform one.
Similarly the above configuration cannot be smoothly deformed to
$\Phi~=~I$ which is also a solution. The above configuration corresponds to a winding number one
map from the physical space $S_F^2$ to the vacuum manifold which is also 
$S_F^2$. All topologically stable configurations in the continuum limit,
can be characterised by the second homotopy group $\Pi_2(S^2)$.
For a discussion on topological classification of the maps $S_F^2\longrightarrow S_F^2$
see \cite{trghari}.
To study the net effect of topological nature of the background
configuration and non-locality on fluctuations, we consider only the
winding number one configuration which is given in Eq.(\ref{winding}).
As mentioned our plan is to compute these
fluctuations numerically. Even before computing the fluctuations
one can make some general remarks about the behaviour of the
fluctuations \cite{digal2}. The effect of nonlocality basically
provides a non-zero mass $O({\alpha}/{N})$ to the Goldstone mode fluctuations. This puts
an infrared cut-off for the fluctutations. From our previous study
\cite{digal2} it seems that this mass/cutoff is mode dependent, as only
higher modes of the condensate survived the fluctuations. As we will see
from our results the combined effect of topology and nonlocality, the infrared
cut-off  drastically  reduce the contribution of the fluctuations.
Before presenting details of numerics let us also consider scalar field on fuzzy cylinder.
\subsection{The Action on the fuzzy cylinder}
Define $ \widetilde{Tr} O~=~Tr(P O P) $ where $P$ is the projection operator defined in
[Eq: No]
This trace $ \widetilde{Tr} $ is equivalent to integrating over 
the whole cylinder in the continuum limit. We also need the derivatives $ \pd_\phi $ and $ \pd_Z $. They are:
\beq
\pd_\phi \Phi~&=&~[Z,\Phi] \label{Zgen} \\
\pd_Z\Phi~&=&~ e^{-i\phi}[e^{i\phi},\Phi]
\eeq
Then, apart from $J$-dependent normalisation factors, the
action can be chosen as:
\be 
S~=~\widetilde{Tr}\left(|~[Z,\Phi]~|^2 ~+~ |~e^{-i\phi}[e^{i\phi},\Phi]~|^2~+~
V(\Phi)\right) \label{Sgen}
\ee
where $V(\Phi)$ is the potential which can be taken to be of the form,
\be
V(\Phi)=\mu\Phi^2+c\Phi^4 \label{potential}
\ee
for a hermitian field $\Phi$. 

This action has a problem of instability coming from
$\widetilde{Tr}(\Phi^4)~=~ Tr((P\Phi)\Phi^2(\Phi P))$ which cannot contain any
quartic (nor cubic) term for the variable $\Phi_{2J\,2J}$. 
This makes the theory unstable with respect to this variable.
The simplest cure is to insist that this term is not a degree of freedom
of the theory and constrain it to be zero. To keep the set of fields an algebra, 
we set to zero the last row $\Phi_{2J\,i}$ and column $\Phi_{i\,2J}$ 
of the field. As a result the hermitian field
$\Phi$ now only has $(2J-1)^2$ degrees of freedom, and $\Phi=P\Phi P$.

\subsection{Dimensional reduction}
With this new choice of the field  the action becomes 
\beq
S & = & Tr\left(~P~|~[Z,P\Phi P]~|^2P~+~P~|~e^{-i\phi}[e^{i\phi},P\Phi P]~|
^2~P+~V(\Phi)\right) \nonumber \\ 
& = & Tr\left(|~[PZP,\Phi]~|^2~+|~[Pe^{i\phi}P,\Phi]~|^2+~V(\Phi)\right)
\eeq
which can be rewritten simply as the action for a hermitian matrix in a
$(2J-1)\times(2J-1)$ matrix algebra of reduced dimension:
\be S~=~Tr_{J'}\left(|~[\tilde{Z},\Phi]~|^2~+|~[\widetilde{e^{i\phi}},
\Phi]~|^2+~V(\Phi)\right) \label{reducedS}
\ee
where $J'=J-1/2$ is the reduced angular momentum, while $\widetilde{e^{i
\phi}}$ and $\tilde{Z}$ are the matrices obtained from $e^{i\phi}$ and $Z$
by removing the last line and column. For $\widetilde{e^{i
\phi}}$, it is equivalent to setting $e^{i\b}\rightarrow 0$ in its
$2J'$-dimensional expression (\ref{newangle}). As for $\tilde{Z}$, it is
therefore the $2J'\times2J'$ diagonal matrix obtained from $Z$ by
removing its top eigenvalue $J-1/2$:
\beq
\tilde{Z} & = & \mbox{Diag}(-J+1/2,-J+3/2,\cdots,J-3/2)=\mbox{Diag}
(-J',-J'+1,\cdots,J'-1) \nonumber \\ \Rightarrow \tilde{Z}_{ij} & = & (-J'-1+i)
\delta_{i,j}.
\eeq
Note that $Z$ and $\tilde{Z}$ are defined by their commutation
relation (\ref{angle}) and only appear in the action through
$\partial_\phi$ as a commutator. As a result, they are only defined up
to a translation by a matrix proportionnal to the unit, and thus
$\tilde{Z}=\mbox{Diag}(1,\cdots,2J')$ is another possible choice.

Although $\widetilde{e^{i\phi}}$ is not unitary, the equation
\begin{eqnarray*}
<m'|[\tilde{Z},\widetilde{e^{i\phi}}]|m> & = & (m'-m)\delta_{m',m+1}=\delta 
_{m',m+1}=<m'|\widetilde{e^{i\phi}}|m>\mbox{ if } m<2J'\\
& = & 0=<m'|[\tilde{Z},\widetilde{e^{i\phi}}]|2J'>\mbox{ if }
m=2J'
\end{eqnarray*} 
shows that $\tilde{Z}$ and $\widetilde{e^{i\phi}}$ do satisfy the
commutation relation (\ref{angle}) for $\alpha=1$

The cylinder is also parametrised by its radius $r$. According to
(\ref{qntmalg}), $\hat{r}$ commutes with both $Z$ and
$e^{i\phi}$. It can therefore be considered as a pure number in the
non-commutative cylinder algebra.

The radius will appear as a simple scaling in the action. The volume
of the cylinder $Tr(1)$ depends linearly on $r$, so the action should
have an overall scale of $r$. The derivative along the axis
$\partial_Z$ does not scale with $r$, whereas the angular derivative
$\partial_\phi$ scales like $1/r$.

As a result, the action on a fuzzy cylinder of radius $r$ is given by:
\beq
S&=&r\,\widetilde{Tr}\left(\frac{1}{r^2}|~[Z,\Phi]~|^2 ~+~ |~e^{-i\phi}[e^{i\phi},\Phi]~|^2~+~V(\Phi)\right)\nonumber \\
&=& r~Tr_{J'}\left(\frac{1}{r^2}|~[\tilde{Z},\Phi]~|^2~+|~[\widetilde{e^{i\phi}},
\Phi]~|^2+~V(\Phi)\right) \label{actionr}
\eeq
\subsection{Spectrum of the Laplacian}
The spectrum of the Laplacian can be obtained from:
\beq
\cL^2\Phi & = & [\tilde{Z},[\tilde{Z},\Phi]]+\frac{1}{2}([\widetilde{e
^{i\phi}}^\dag,[\widetilde{e^{i\phi}},\Phi]]+[\widetilde{e
^{i\phi}},[\widetilde{e^{i\phi}}^\dag,\Phi]]) \nonumber \\ & = & \mathcal{L}^2_Z
\Phi+\frac{1}{2}(\mathcal{L}_{_-}\mathcal{L}_{_+}\Phi+\mathcal{L}_{_+}
\mathcal{L}_{_-}\Phi) \label{Laplacian} \\
& = & \mathcal{L}^2_Z\Phi+\mathcal{L}_{_-}\mathcal{L}_{_+}\Phi+
\frac{1}{2}[[\widetilde{e^{i\phi}},\widetilde{e^{i\phi}}^\dag],
\Phi]=\mathcal{L}^2_Z\Phi+\mathcal{L}_{_-}\mathcal{L}_{_+}\Phi+\frac{1}{2}
[\mbox{Diag}(-1,0,\cdots,0,1),\Phi]\nonumber\eeq

The eigenmatrix equation for the
Laplacian reads simply
$$\mathcal{L}^2\Phi_m(\vec{d})=\lambda\Phi_m(\vec{d})=\Phi_m(\lambda\vec{d})
\Leftrightarrow
M_m\vec{d}=\lambda\vec{d}$$
where
\be
\Phi=\Phi_m(\vec{d})+\Phi_m^\dag(\vec{d}),\mbox{ with }
\Phi_m(\vec{d})=\sum_{i=1}^{2J'-m} d_i\,|i><i+m|,\ 0\leq m\leq
2J'-1 \label{evansatz}
\ee
for the eigenmatrices, and the vector $\vec{d}=(d_i)_{1\leq i\leq
2J'-m}$ is the unknown to be determined by the eigenmatrix equations.

For the Laplacian on a cylinder of radius $r$, we get
\be
M_0=\left(\begin{array}{cccccc} 1&-1&0&&\\-1&2&-1&&(0)\\
&\ddots&\ddots&\ddots&\\(0)&&\ddots&2&-1\\&&&-1&1\end{array}\right),\
M_m=\frac{m^2}{r^2}+\left(\begin{array}{cccccc} 3/2&-1&0&&\\-1&2&-1&&(0)\\
&\ddots&\ddots&\ddots&\\(0)&&\ddots&2&-1\\&&&-1&3/2\end{array}\right),\ m
\not=0
\ee
These matrices are similar to the ones obtained for the Laplacian on a
one-dimensional lattice and can actually be diagonalised without much difficulty,
taking good care to remember that $M_m$ is a matrix of dimension
$2J'-m$.

\paragraph*{\underline{Spectrum of $M_0$}}
The eigenvalues of $M_0$ are:
\be
\lambda_0^k=4\sin^2(k\pi/4J'),\ 0<k\leq 2J'.
\ee
\paragraph*{\underline{Spectrum of $M_m$, $m\not=0$}}
Reparametrizing the eigenvalues as
\be
\lambda=2-2\cos(\theta)=4\sin^2(\t/2),\label{ev}
\ee
We can solve for eigenvalues for any $N$ numerically. However
for large matrices $N\gg 1$, it is possible to find approximate ones:
\begin{itemize}
\item For $\theta\ll \pi$,  or $k\ll N$, $\tan((\pi-\t)/2)\sim 2/\t\gg
  1$. Therefore,  $N\t=k\pi+\pi/2-\rho_k$ with $\rho_k\ll 1$. The
  equation then becomes:
$$\frac{1}{\rho_k}\simeq \frac{2N}{3(k\pi+\pi/2-\rho_k)}\Leftrightarrow
\rho_k\simeq\frac{3\pi}{2N}(k+1/2).$$
\item For $\theta\simeq \pi$, or $k\sim N$,
  $\tan((\pi-\t)/2)\sim (\pi-\t)/2\ll 1$, and therefore,
  $N\t=k\pi+\rho_k$ with $\rho_k\ll 1$. The equation then becomes:
$$\rho_k\simeq\frac{N\pi-k\pi-\rho_k}{6N}\Leftrightarrow
\rho_k\simeq \frac{N-k}{6N}\pi$$
which is a small number, as expected, since $k\sim N$.
\end{itemize}

\section{Numerical Simulations and Results}
Effects of the fluctuations beyond mean field are computed from
the partition function, which in the path integral approach
is given by,

\begin{equation}
{\cal Z}\; \propto\; \int D\Phi e^{-S(\Phi)}.
\end{equation}

The standard numerical methods adopted for this
integration are Monte Carlo simulations. 
\subsection{The numerical scheme: pseudo heatbath}
In the Monte Carlo algorithms,
one generates an ``almost'' random sequence of $\Phi$ matrices by
successively updating elements of $\Phi$ taking into account the measure
and the exponential in the integral above. This sequence
of $\Phi$ is then used as an ensemble for calculating averages of various
observables. For a good ensemble the auto correlation
between the configurations in the sequence must be really small. Though this
auto correlation can be reduced by using some over relaxation programme
\cite{panero}, it is greatly reduced, however, when
``heatbath/pseudo-heatbath'' type of algorithms are used. This
method is very much common in the non-perturbative study of $\Phi^4$ theories
in conventional lattice simulations. It gives better sampling and is efficient
at least for smaller $\lambda$ values. This is why we make use of
``pseudo-heatbath'' technique \cite{digal1,digal2}.
\subsection{Topological stability and O(3) model}
In our simulations, for each choice of parameters, we choose an initial
configuration given by Eq. (\ref{winding}). Fluctuations around this
configuration are then generated by the above updating method.
Since this configuration is a variational solution to minimising the
classical action, it will thermalise as we update/include the thermal
fluctuations. Once the initial configuration is thermalised we compute
the observable M. We make measurements
after every 10 updates of the entire matrix. We also use over-relaxation to
reduce the auto correlation of the configurations generated in the
Monte-Carlo history.

In a numerical simulation, the condensate  will not maintain
its exact form as in Eq. (\ref{winding}) along the Monte Carlo history. The configuration
can evolve into different random $SU(2)$ rotated configurations of Eq.(\ref{winding})
as we keep updating it. To overcome this, one needs to rotate the configuration at
each step of the Monte Carlo history so that the configuration takes the
form of Eq. (\ref{winding}). But this is a difficult and time consuming task.
On the other hand
one can have an observable made of $\Phi_i'$s which is invariant under the
$SU(2)$ rotations, e.g basis independent. For this purpose we define the
following observable,
\begin{equation}
A_{ij} = \frac{1}{N^2} Tr(L_i\Phi_j), M = \sqrt{A^\dag A}
\end{equation}
$M$ projects out the $l=1$ angular momentum mode. Note that the initial
configuration in Eq. (\ref{winding}) projects out only the $l=1$ mode. Analysing the
statistical behavior of $M$ will give us a definite conclusion about the stability of the
initial configuration. We mention here that
$Tr(\sum_i \Phi_i^2)$ is also an $SU(2)$ invariant. But the information on the
amplitudes of different $l$ modes gets lost in this form. Also comparatively
the observable $M$ may serve as an order parameter in the case of any
phase transition of the hedgehog configuration to $\Phi_i = 0$ at high
temperatures. We mention here that $l=1$ is the lowest possible stable mode,
as $l=0$ mode will be unstable. One can consider configurations with higher
winding, instead of Eq.(\ref{winding}), however we expect them to be more stable than
the $l=1$ condensate. This is because the infrared cut off will rise
with higher winding configurations.

For practical reasons, the size of the matrix $N$, in other words size of
the resolution scale is finite. So there are usually finite volume $(R,N)$
effects. So a non-vanishing condensate $\Phi_i$ does not mean
there is SSB. One needs to define suitable observable dependent on $\Phi_i$
which should scale with $(R,N)$ appropriately in the thermodynamic limit
$(N \rightarrow \infty,R \rightarrow \infty)$ to conclude anything. Now there are
two possible thermodynamic limits. If in the thermodynamic limit the ratio
${R^2}/{N}$ does not vanish then the space is described by a non-commutative algebra.
This limit  is of interest to us, as  we expect that the CMW theorem will hold good
in the commutative thermodynamic limit.

\begin{figure}[hbt!]
\begin{center}
{ \label{fig1a:subfig-1}\includegraphics[scale=0.59]{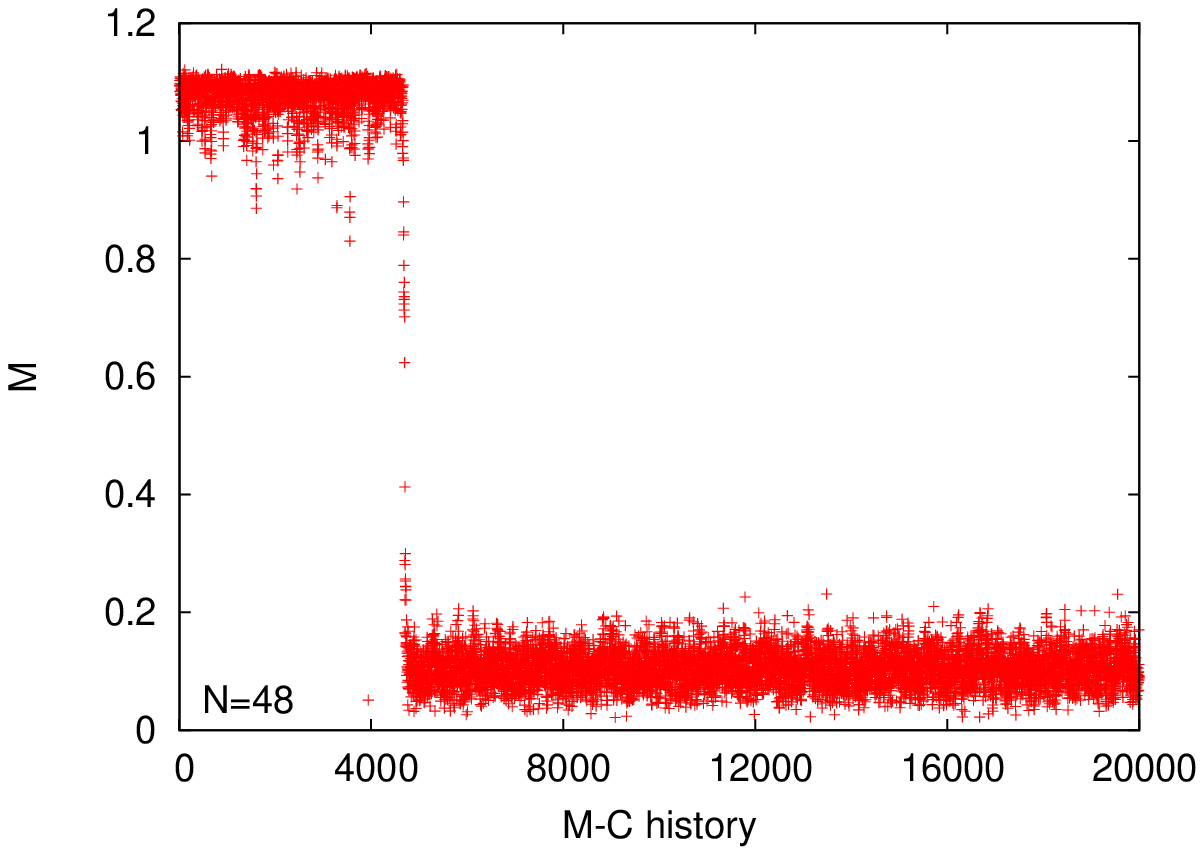}}
\caption{\label{fig1a}Monte Carlo history for N=48}
{ \label{fig1b:subfig-2}\includegraphics[scale=0.59]{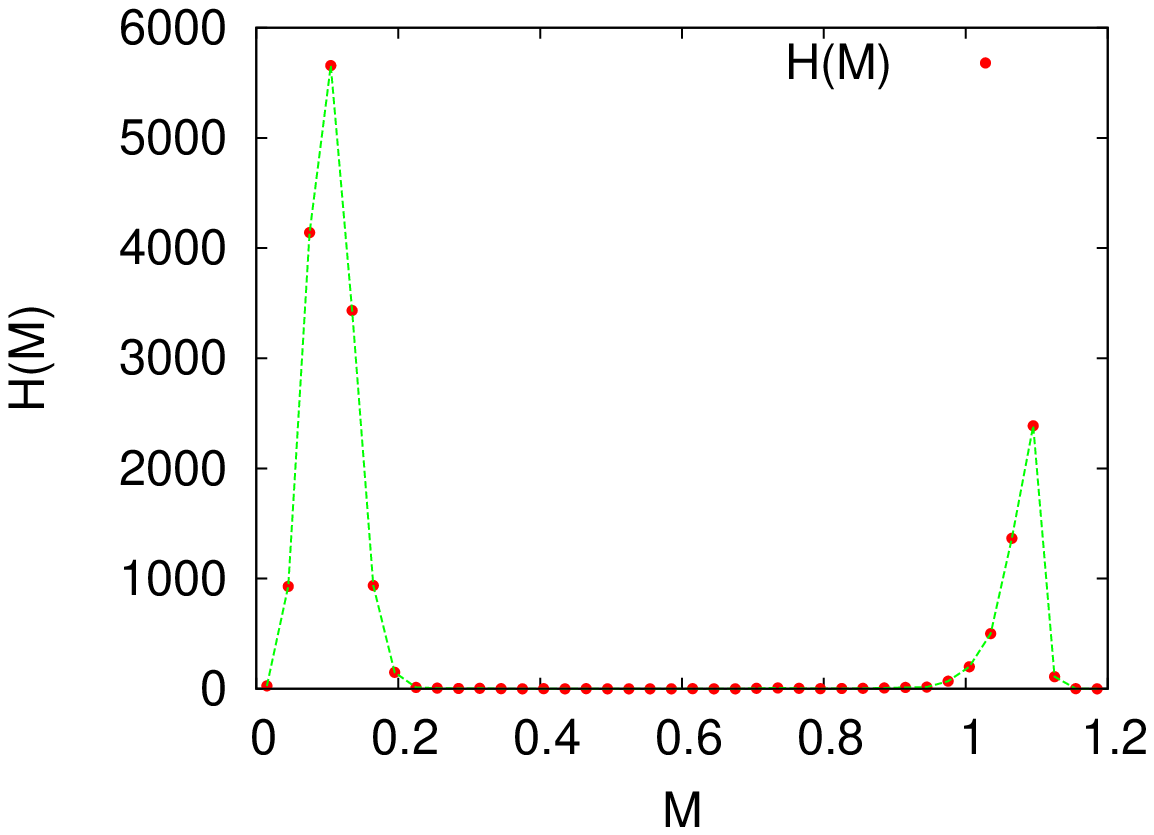}}
\caption{\label{fig1b}Histogram H(M) for N=48}
\end{center}
\end{figure}
\subsubsection{Results and discussions}
In our calculations we fix ${R^2}/{N} = 10$, $r = -8$. For simplicity
we take $\lambda_1 = 0.25$. With this choice of parameters
we do our simulations for five different sizes of the $\Phi$
matrices, $N=48, 56, 64, 78, 96$. Fig.\ref{fig1a} gives a typical Monte-Carlo
history of our simulations for $N=48$.

In Fig.\ref{fig1a}  $M$ fluctuates around a value close to the initial value.
Then $M$ suddenly jumps to a small value and settles down. A histogram $H(M)$ of
$M$ clearly shows two peaks $M$ as seen in Fig.\ref{fig1b}. The peak on the
left has large $l=0$ and small $l=1$ component. The peak at higher
value of $M$ has large $l=1$ component and small $l=0$. This peak is
close to the value of the initial configuration. So in this state
fluctuations modify the initial configuration slightly and retain its
topological nature. In our Monte-Carlo history we observed the $l=1$ state
decaying to $l=0$ state but not vice-versa. This implies that due to
finite volume effects, the uniform condensate is more stable than our
initial hedgehog configuration for this case of $N=48$.

\begin{figure}[hbt!]
\begin{center}
{ \label{fig2a:subfig-1}\includegraphics[scale=0.59]{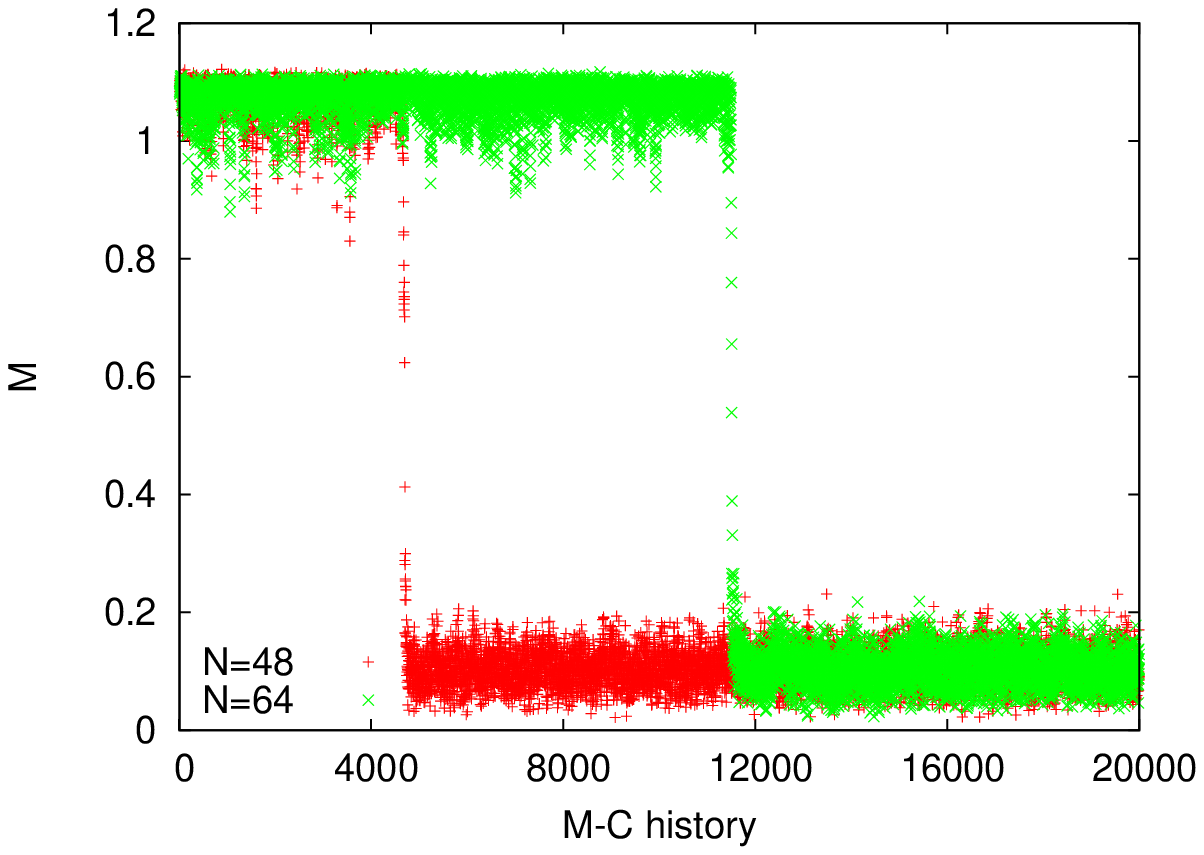}}
\caption{\label{fig2a}M-C history for fixed $R^2$}
{ \label{fig2b:subfig-2}\includegraphics[scale=0.59]{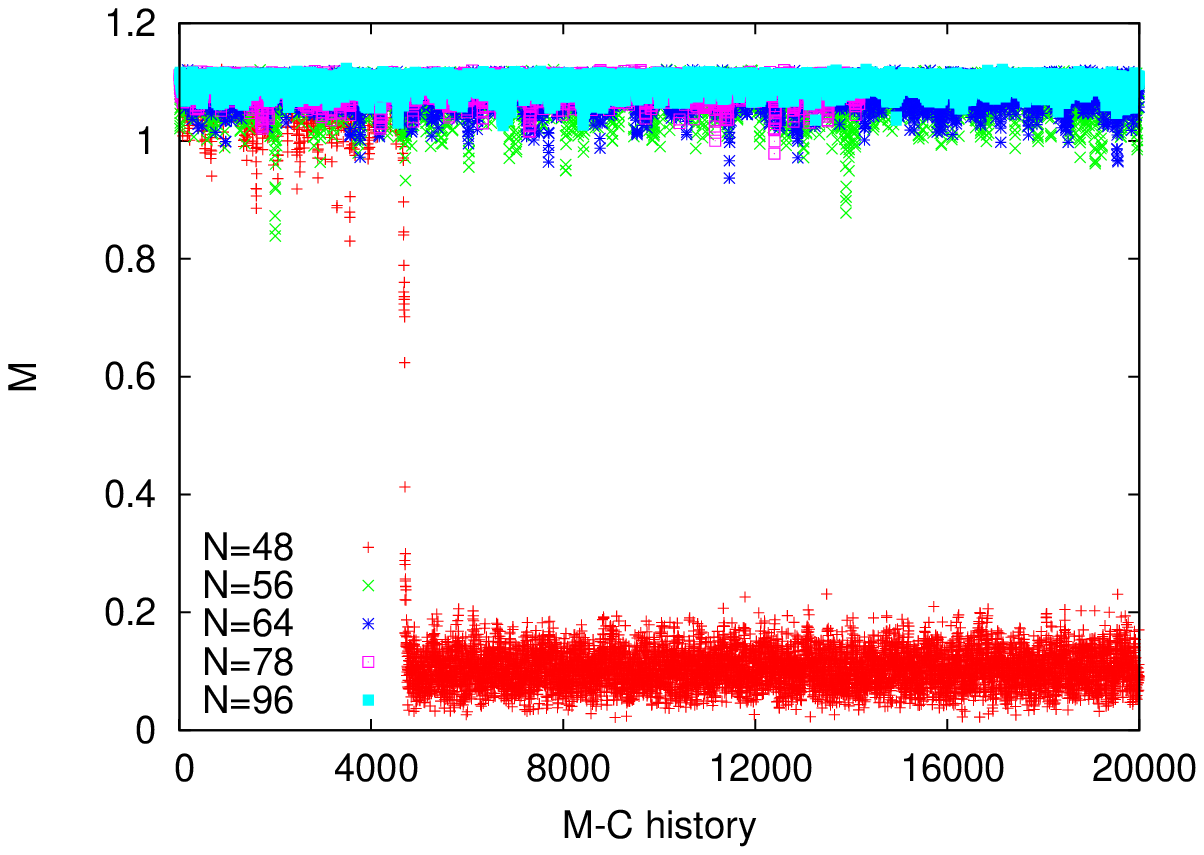}}
\caption{\label{fig2b}M-C history for fixed $R^2/N$}
\end{center}
\end{figure}

To study the stability of the $l=1$ configuration we considered both the
commutative and non-commutative limit. For the commutative limit we fixed
$R^2$ and considered higher values of $N$. We did not observe any
change in the distribution of $M$ in the $l=1$ state. The average value,
and the fluctuations of $M$ remain almost the same as we go from
$N=48 \rightarrow 64$, as can be seen in Fig.\ref{fig2a}. As for $N=48$ the $l=1$
configuration also decays for $N=64$. This result suggests that the
$l=1$ topological configuration is not stable in the commutative
continuum limit as expected.

\begin{figure}[hbt!]
\begin{center}
{ \label{fig3a:subfig-1}\includegraphics[scale=0.59]{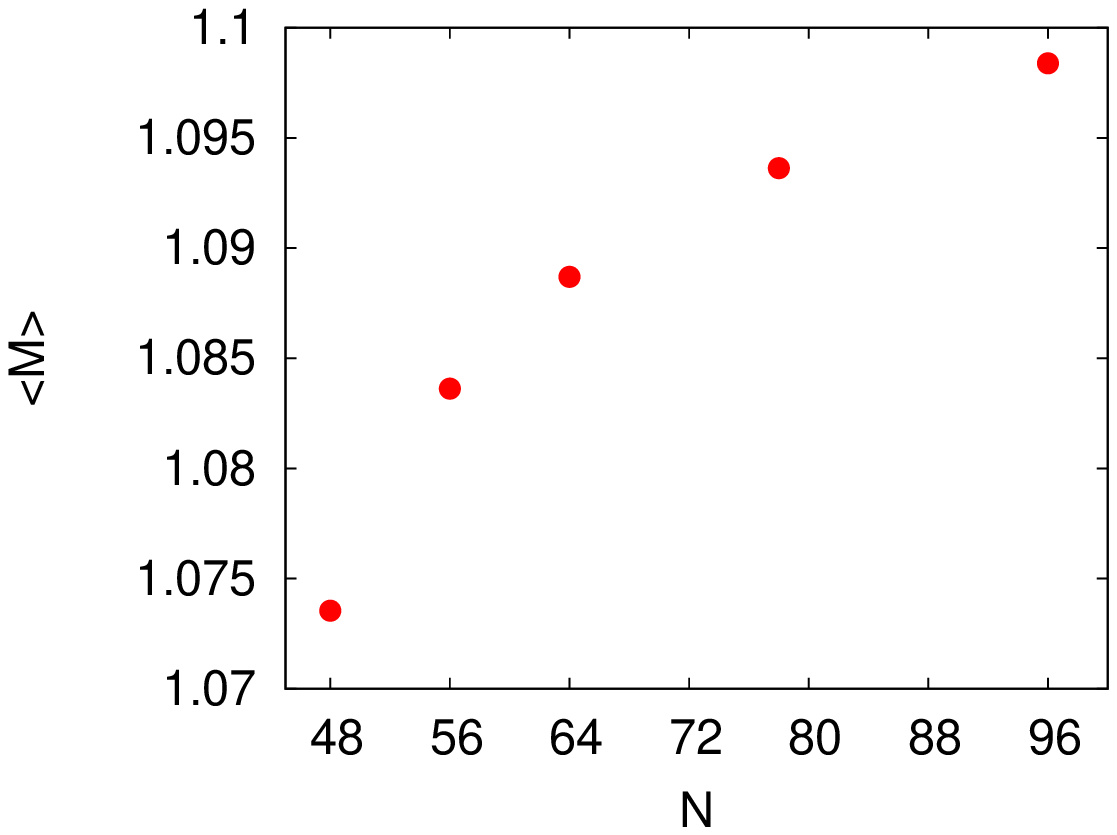}}
\caption{\label{fig3a}$\left<M\right>$  vs  $N$}
{ \label{fig3b:subfig-2}\includegraphics[scale=0.59]{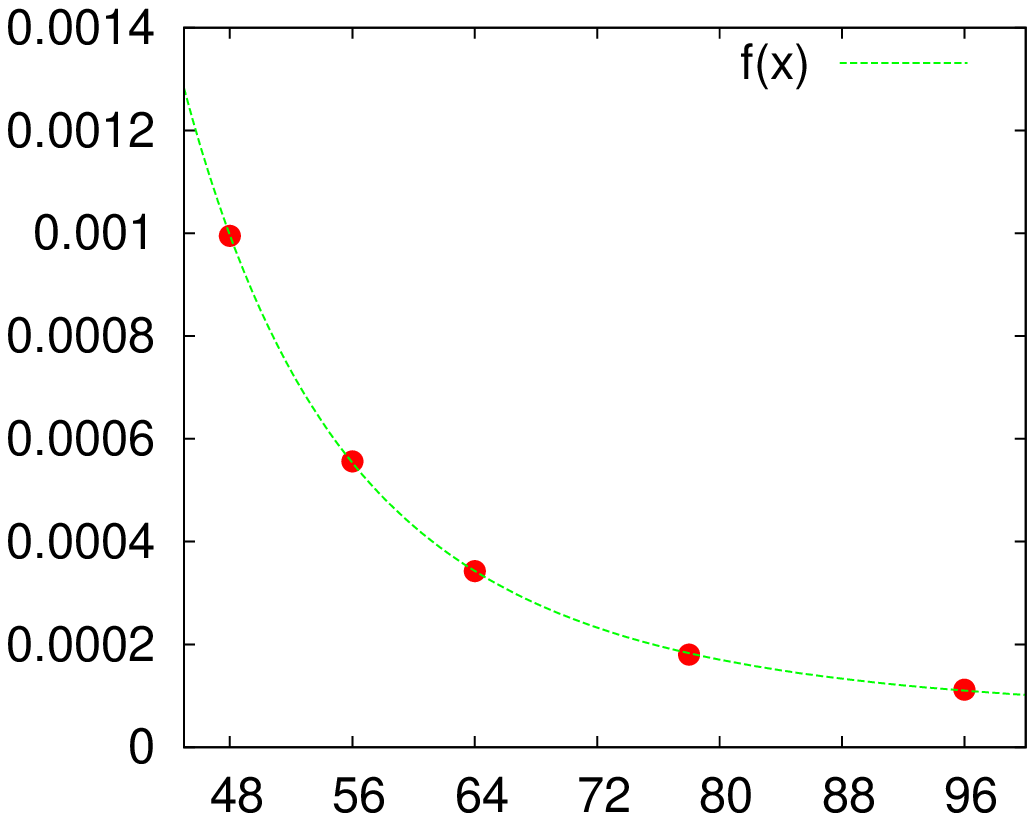}}
\caption{\label{fig3b}$\chi$ vs $N$}
\end{center}
\end{figure}

There is a complete change in the behavior as we consider the
non-commutative limit, i.e fixed ${R^2}/{N}$ as we
increase $N$. Except for the lowest $N=48$ the $l=1$ state did not decay
during the entire run for higher $N$. In Fig.\ref{fig2b} we show the Monte-Carlo
history of $N=48,64,96$. Unlike the commutative limit, the fluctuations of
$M$ decrease with $N$. In Fig.\ref{fig3a}, we
give the  average value of $M$ as a function of $N$.
The average value of $M$ increases slightly
with $N$, with the variation decreasing with $N$. This suggests $M$ will
reach a finite value in the continuum limit. We also compute the fluctuations
of $M$ to see any possible scaling with the cut-off $N$.
In Fig.\ref{fig3b}, we show $\chi~=~\left< M^2 \right > - \left< M \right> ^2$
in the $l=1$ state. The solid curve represents a fit, $f(N) \sim N^\alpha$
with $\alpha\sim -4.$. This clearly suggest that the $l=1$ state is stable
in the $N \rightarrow \infty$ leading to spontaneous
breaking of the $O(3)$ symmetry.

We also mention here that one can start with an initial uniform $l=0$
configuration and consider fluctuations. We expect that the results be similar
to that in ref.\cite{digal2}. In ref.\cite{digal2} it was
found that only the highest mode $l = {(N-1)}/{2}$ condenses. The fact
that we find the $l=1$ mode stable clearly shows that the topological nature
of the initial configuration complements the effect of non-locality. These two
effects drastically reduce the fluctuations.
\subsection{Fuzzy blackholes: NC cylinder}
The model defined by the action Eqs.(\ref{potential},\ref{actionr})
which we want to simulate has three parameters $(\mu,c,r)$ plus the
matrix size $J$.  The goal is to explore the parameter space for
various phases of $\Phi$.  The simulations are carried out using the
"pseudo-heat bath" Monte-Carlo (MC) algorithm \cite{digal1,digal2} to
reduce the auto-correlation along the MC history.

The field should also be allowed to explore the whole phase space and
not remain trapped in local minima. To this end, an over-relaxation
method, first suggested in \cite{panero}, is also used. Let us
introduce $S_\Phi(\Phi_{ij})$ the dependence of the action on the
field entry $\Phi_{ij}$ when the field takes the value $\Phi$. It is a
fourth degree polynomial. Therefore the equation
$S_\Phi(\Phi_{ij})=S_\Phi(\Phi_{ij}=a)$, which has an obvious solution
$\Phi_{ij}=a$, can be factorised into a degree three polynomial which
{\it always} admits at least one real solution. The overrelaxation
method consists in replacing the field entry $\Phi_{ij}=a$ by one of
these real solutions, thereby moving the field in a different region
of the phase space.
A crosscheck is also used to verify that the field probability
distribution of our Monte-Carlo runs are consistent. Let us split the
terms in the action according to their scalings
$$S(\phi)=S_2(\phi)+S_4(\phi)\mbox{ with } S_i(\lambda\phi)=\lambda^i
S_i(\phi).$$
Then one can define a modified partition function
\beq
Z(\lambda)=\int[\diff\phi]e^{-S(\lambda\phi)} & = & \int[\diff\phi]
e^{-\lambda^2S_2(\phi)-\lambda^4S_4(\phi)} \label{Z1} \\
& = & \lambda^{-N}\int[\diff\psi]e^{-S(\psi)},\ \psi=\lambda\phi,
\label{Z2}
\eeq
where $N$ is the number of degrees of freedom in the field $\phi$
which appear in the integration.
Evaluating
\begin{eqnarray*}
\left.\frac{\partial \ln(Z)}{\partial\lambda}\right|_{\lambda=1} & = &
-2<S_2>-4<S_4> \mbox{ from (\protect\ref{Z1})} \\
& = & -N\mbox{ from (\protect\ref{Z2})}
\end{eqnarray*}
yields the check originally due to Denjoe O'Connor \cite{denjoe1}.
\be
<S_2>+2<S_4>=N/2
\label{Did}\ee
In all simulations, this identity (\ref{Did}) is always satisfied to better than $1\%$ relative error.

\subsubsection{The phase structure}
The temperature ($T$) is regulated by varying the parameter
$\mu$. \begin{itemize}
\item $\mu\ll 1$ corresponds to low temperatures when the fluctuations are small. In
this case, the minimum of $S$ gives the most probable configuration of
the phase. In Eq. (\ref{actionr}), it is possible to minimise
the action by minimising separately the kinetic term, so that $\Phi
\propto {\bf 1}$, and the potential term so that
$\Phi=\sqrt{-\mu/2c}\,{\bf 1}$, and this phase is therefore known as the
uniform phase.
\item At high temperatures, $ \mu\gg 1$, the thermal
fluctuations lead the system to the disorder phase $\Phi \sim 0$.
\item At intermediate temperatures, the competition between the action and the
fluctuations give rise to new phases called the non-uniform or stripe
phases. These new phases are specific to non-commutative
spaces. Various numerical studies have confirmed the existence of
these phases \cite{xavier,denjoe,panero,digal1,digal2,bietenholz,
  medina,ambjorn} on the fuzzy sphere.
A non-commutative cylinder will also exhibit the non-uniform
phases. However, due to the non-trivial topology of the cylinder (the
first homotopy group being non-trivial), one can have a more complex
phase structure described below.
\end{itemize}
For example there can be stripes going
around the cylinder, or parallel to its axis. These two phases can be
distinguished by their overlap with the operators $Z$, ${e^{i\phi}}$,
and ${e^{i\phi}}^\dag$
respectively. Stripes going around the cylinder will have non-zero
overlap with the operator $Z$. While a configuration of stripes along
the axis will have overlap with ${e^{i\phi}}$ and
${e^{i\phi}}^\dag$. We present our results in the following subsection.

\subsubsection{Example numerical runs}
For a given choice of $N=7$, $c = 0.36$, and $r=1$ the simulations are
done for various values of $\mu$. The various phases discussed above can
be characterised by the observables $m_u=Tr(\Phi)$, $m_z=Tr(\Phi Z)$,
$m_x = Tr(\Phi e^{i\phi})$. A finite $m_u$ with $(m_z,m_x) \sim 0$
characterizes the uniform phase. On the other hand, $(m_u,m_x)\sim 0$
with non-zero $m_z$ characterizes stripes going around the
cylinder. Stripes along the cylinder charactersied by $(m_u,m_z)\sim
0$ with non-zero $m_x$.

\begin{figure}[hbt!]
\begin{center}
{ \label{uprun:subfig-1}\includegraphics[scale=0.59]{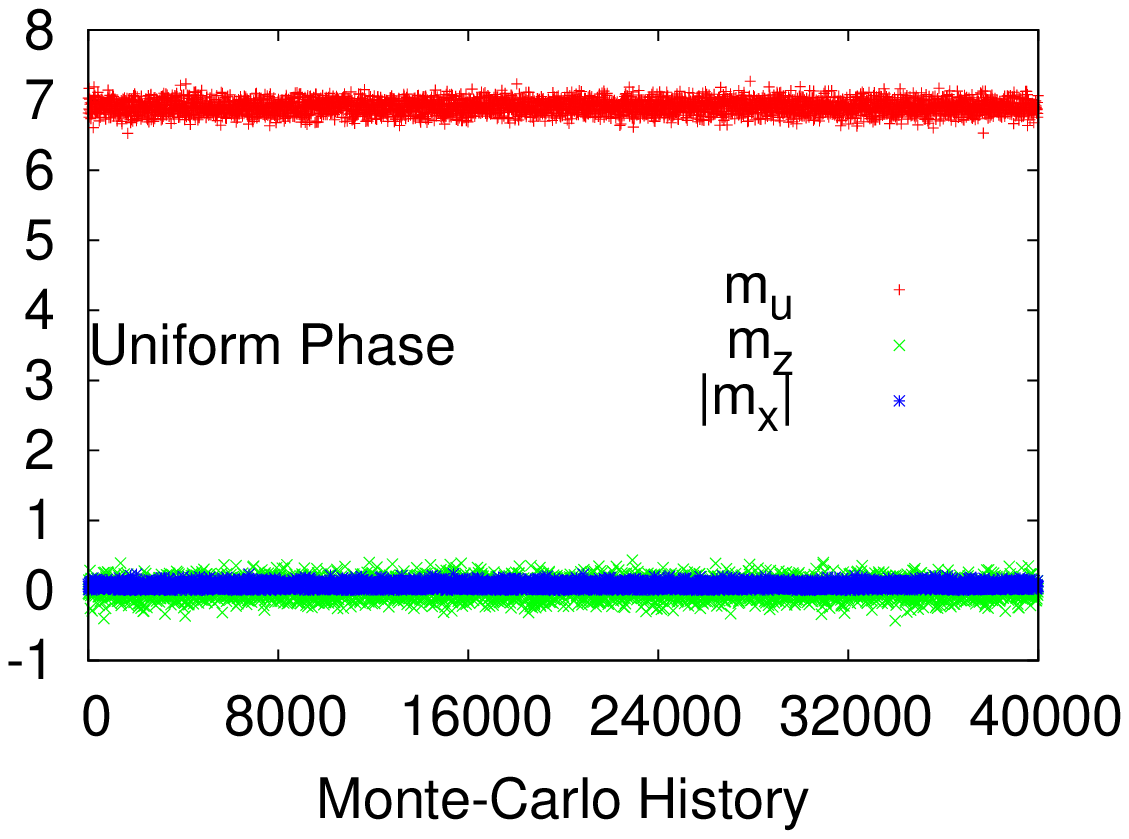}}
\caption{\label{uprun}$m_{u,z,x}$ vs MC history $N=7, \mu = -35.1, c = .36$}
{ \label{histo:subfig-2}\includegraphics[scale=0.59]{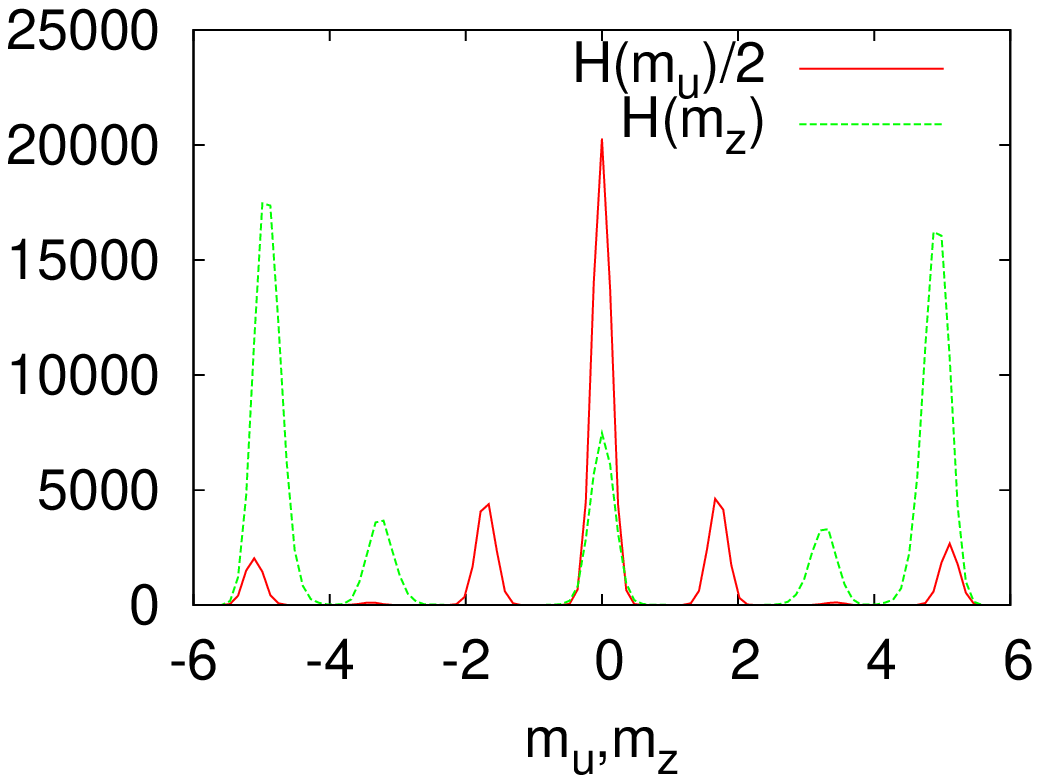}}
\caption{\label{histo}Histogram of $H(m_u), H(m_z): N=7,\mu = -20$}
\end{center}
\end{figure}

For $\mu = -35.1$, the data of a run are shown on Fig.\ref{uprun},
and, as expected, we observe the uniform phase.

For $ \mu = -20.0$, we observed the phase with stripes going around the
cylinder. This is verified on the histogram of the observed values of
$m_u,m_z$ plotted in Fig.\ref{histo}. It is clear from the figure
that the average value of $m_z$ is finite while the average value of
$m_u$ is vanishingly small.

\begin{figure}[hbt!]
\begin{center}
{ \label{disrun:subfig-1}\includegraphics[scale=0.59]{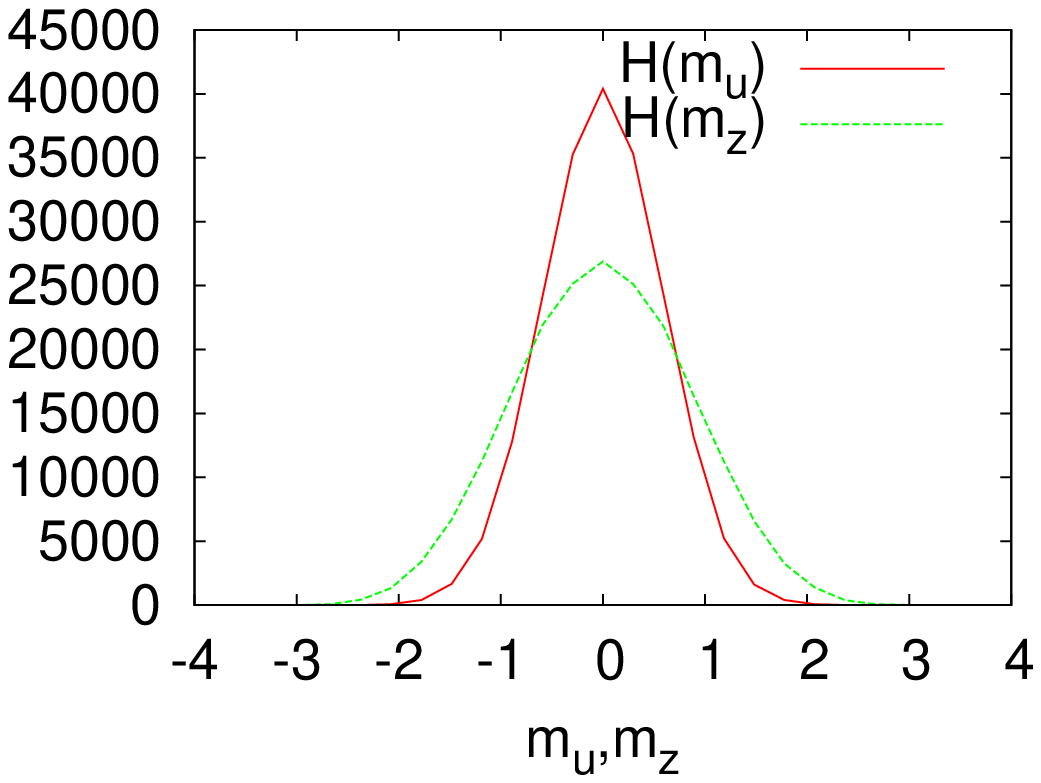}}
\caption{\label{disrun}Histogram of $H(m_u),H(m_z): N=7= 0.36$}
{ \label{stripesrun:subfig-2}\includegraphics[scale=0.59]{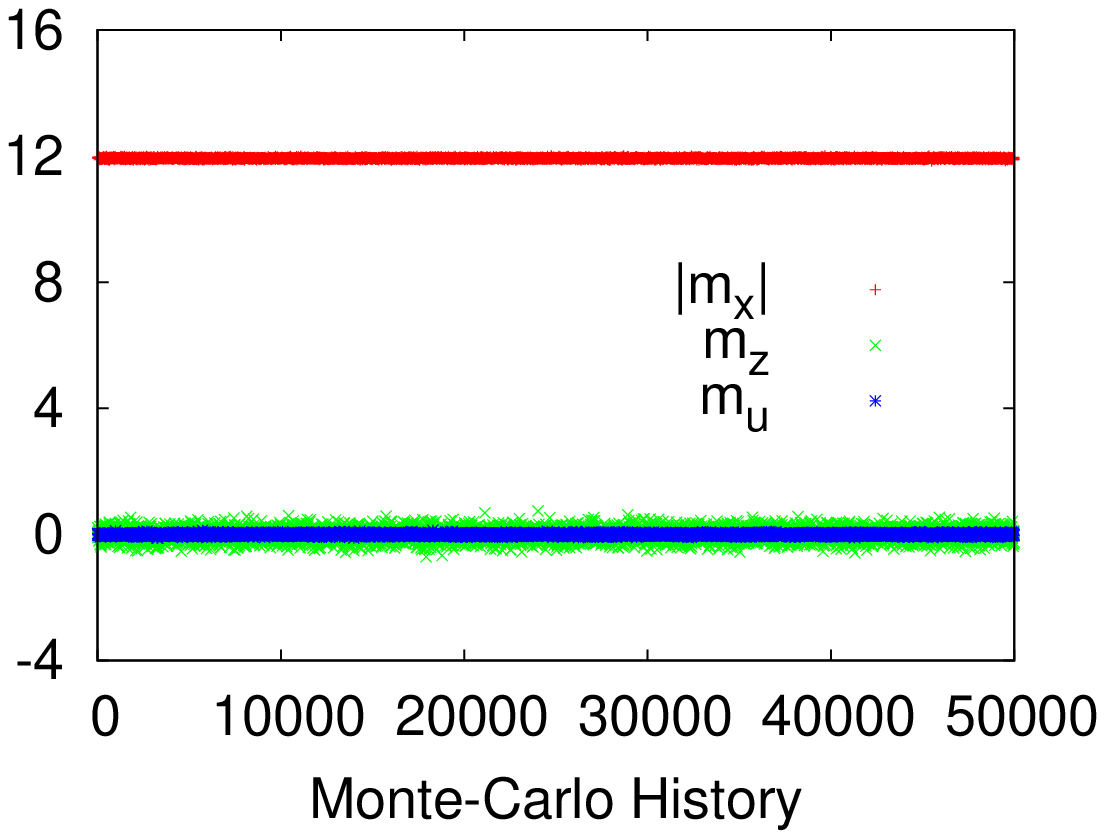}}
\caption{\label{stripesrun}$m_{u,z,x}$ vs MC history $N=7, \mu= -3.6\times 10^3, c=36, r=.01$}
\end{center}
\end{figure}

Fig.\ref{disrun} shows the system in the disorder phase where
$m_u,m_z,m_x$ all fluctuate around zero.
We did not observe the phase with stripes going along the cylinder as a
ground state for any choice of $\mu$ for $r\sim 1$. One can expect to
observe this state for very small $r$ when the second term which suppresses this 
state is made subdominant. For a very small radius
$r=0.01$, $c=36.$ and $\mu= -3.6 \times 10^3$, this phase appears
as meta-stable in Fig.\ref{stripesrun}. This phase is stable w.r.t small
fluctuations. Only large fluctuations, which occur less frequently,
can destroy such a state.

\section{Conclusions}

In this analysis we have shown that topologically non-trivial
configurations on the fuzzy sphere
avoid the CMW theorem much more dramtically than the non-topological symmetry breaking
\cite{digaltrg}.
The mass gap or the infrared cut-off in this case is
large enough to render the fluctuations of the Goldstone modes finite. On the
other hand for non-topological condensates the Goldstone modes are large
enough to destroy almost all the modes except the few highest modes.
We have presented the simulations wherein the cubic Chern-Simons (CS) term is absent
in the action Eq (\ref{action}).
The Chern-Simons term allows topological solitons even when the
quadratic mass term is positive upto some value.
Interestingly with CS term the configuration $\phi_i ~=~ \alpha~L_i$
is preferred over symmetric solution.
On the otherhand, it is not expected
to alter the picture of topological stability of the solutions. This term plays an important
role in the emergent geometry in NC fuzzy spaces \cite{steinacker,ydri}.
What we find here in the simulations is that even in the absence of CS term, emergent fuzzy spaces can be stable.
The stabilty of higher dimensional fuzzy spaces like $CP_F^2$ are of significance in this
context \cite{dolan}. The implications of this stability for extra-dimensional fuzzy spaces will
be considered later.

We have also considered a finite dimensional representations of
the noncommutative cylinder algebra, which make it fuzzy. We study
scalar field theory in the background of this algebra both analytically
and using numerical simulations.

In the numerical simulations of scalar field with a generic potential 
we find, as expected in noncommutative cylinder, novel stripe phases
breaking translational symmetry.
But they have some differences with the usual stripes on 
Moyal spacetimes. These are also stable due to topological features arising 
in this fuzzy geometry. 
   
It is well known that a large class of noncommutative black holes are
described by a noncommutative cylinder algebra. The fuzzy cylinder algebra
derived from it can therefore be used to define a fuzzy black hole. From
general considerations\cite{dop2}  we know that such
black holes can arise at the Planck scale. Our results provide a first
glimpse about the phase structure of a quantum scalar field theory in the
background of a fuzzy black hole at the Planck scale \cite{fuzzybh}.

{\bf Acknowledgment} This work was done as a part of the CEFIPRA/IFCPAR project 4004-1 entitled Fuzzy Approach to Quantum Field Theory and Gravity. The authors gratefully acknowledge the financial assistance from CEFIPRA/IFCPAR which was essential for this work.


\begin{thebibliography}{99}
 
\bibitem{dop2} S. Doplicher, K. Fredenhagen and J. E. Roberts, Commun. Math. Phys. {\bf 172} (1995) 187.

\bibitem{podles} P. Podles, E. Muller,  Rev. Math. Phys. 10 (1998), 511-551  [arXiv:q-alg/9704002v2]

\bibitem{clifford} ``On the hypotheses which lie at the bases of geometry'', Bernhard Riemann, 1854
(from the translation by W K Clifford)

\bibitem{pinzul} A.P. Balachandran, T.R. Govindarajan and B. Ydri,
Mod. Phys. Lett. {\bf A15} (2000) 1279 [hep-th/9911087];\\
A.P. Balachandran, A. Pinzul and B.A. Qureshi, JHEP
{\bf 0512} (2005) 002 [hep-th/0506037].

\bibitem{gubser} S.S. Gubser and S.L. Sondhi, Nucl. Phys. {\bf B605},
395 (2001) [hep-th/0006119].

\bibitem{xavier} X. Martin, JHEP {\bf 0404} (2004) 077 [hep-th/0402230].

\bibitem{denjoe} J. Medina, W. Bietenholz, F. Hofheinz and
D. O'Connor, PoS {\bf LAT2005} (2005) 263 [hep-lat/0509162];\\
F. G. Flores, D. O'Connor and X. Martin, PoS {\bf LAT2005} (2006) 262 [hep-lat/0601012];\\
D. O'Connor and B. Ydri, JHEP {\bf 0611} (2006) 016 [hep-lat/0606013];\\
J. Medina,  Phd. thesis, arXiv: 0801.1284 [hep-th].
\bibitem{bietenholz} W. Bietenholz, F. Hofheinz and J. Nishimura,
Acta. Phys. Polon. {\bf B34} (2003) 4711 [hep-th/0309216];\\
Nucl. Phys. Proc. Suppl. {\bf 129} (2004)  865  [hep-th/0309182].
\bibitem{panero} M. Panero, SIGMA {\bf 2} (2006) 081 [hep-th/0609205];\\
JHEP {\bf 0705}, 082 (2007) [hep-th/0608202].

\bibitem{digal1} C.R. Das, S. Digal and T.R. Govindarajan, Mod. Phys. Lett. {\bf A23}
(2008) 1781.
\bibitem{digal2} C.R. Das, S. Digal and T.R. Govindarajan,
Mod. Phys. Letts {\bf A24} (2009) 2693.

\bibitem{trghari} T R Govindarajan and E Harikumar, Phys. Letts., {\bf
B 602} (2004) 238; A P Balachandran and G Immirzi, Int Jour. Mod. Phys. {\bf
A19} (2004) 5237.

\bibitem{ambjorn} J. Ambjorn and S. Catterall, Phys. Lett. {\bf B549} (2002)
253 [hep-lat/0209106].

\bibitem{medina} J. Medina, W. Bietenholz and D. O'Connor,
JHEP {\bf 0804}, (2008) 041;  arXiv:0712.3366 [hep-th].

\bibitem{wess1} P. Aschieri, C. Blohmann, M. Dimitrijevic, F. Meyer, P. Schupp and J. Wess, Class. Quant. 
Grav. {\bf 22} (2005) 3511.


\bibitem{seckin} A.P. Balachandran, T.R. Govindarajan, K.S. Gupta, S. Kurkcuoglu,  Class. Quant. Grav. {\bf 23} (2006) 5799.

\bibitem{brian1} B.P. Dolan, Kumar S. Gupta and A. Stern, Class. Quant. Grav. {\bf 24} (2007) 1647. 


\bibitem{schupp} P. Schupp and S. Solodukhin, arXiv:0906.2724 [hep-th].

\bibitem{ohl} T. Ohl and A. Schenkel, JHEP {\bf 0910} (2009) 052. 

\bibitem{btz1} M. Banados, C. Teitelboim and J. Zanelli, Phys. Rev. Lett. {\bf 69} (1992) 1849.

\bibitem{btz2} M. Banados, M. Henneaux, C. Teitelboim and J. Zanelli, Phys. Rev. {\bf D  48} (1993) 1506.
\bibitem{petergrosse} H. Grosse and P. Presnajder, Acta Phys. Slovaca {\bf 49} (1999) 185.

\bibitem{L1} J. Lukierski, A. Nowicki, H. Ruegg and V. N. Tolstoy, Phys. Lett.  {\bf B 264} (1991) 331.



\bibitem{L4} J. Lukierski, H. Ruegg and W. J. Zakrzewski, Ann. Phys. {\bf 243} (1995) 90. 

\bibitem{M1} S. Meljanac and M. Stoji\'{c}, Eur. Phys. J. C \textbf{47} (2006) 531.


\bibitem{M3} S. Meljanac, A. Samsarov, M. Stoji\'{c} and K. Gupta, Eur. Phys. J. C \textbf{53} (2008) 295.


\bibitem{luk2} M. Daszkiewicz, J. Lukierski and M. Woronowicz, J. Phys. {\bf A 42} (2009) 355201. 

\bibitem{luk3} J. Lukierski, Rept. Math. Phys. {\bf 64} (2009) 299.

\bibitem{G1} T.~R.~Govindarajan, K.~S.~Gupta, E.~Harikumar, S.~Meljanac and D.~Meljanac, Phys.\ Rev.\  D {\bf 77} (2008) 105010.

\bibitem{G2} T.~R.~Govindarajan, K.~S.~Gupta, E.~Harikumar, S.~Meljanac and D.~Meljanac, Phys.\ Rev.\  D {\bf 80} (2009) 025014.

\bibitem{dice2010} T.R.Govindarajan, Kumar S. Gupta, E. Harikumar and S. Meljanac, Journal of Physics: Conference Series 306 (2011) 012019.

\bibitem{ksms} Kumar S. Gupta, S. Meljanac and A. Samsarov, arXiv:1108.0341 [hep-th]

\bibitem{C2} M. Chaichian, A. Demichev, P. Presnajder and A. Tureanu, Phys. Lett. {\bf B 515} (2001) 426.

\bibitem{paulo} A. P. Balachandran, T. R. Govindarajan, A. G. Martins and P. Teotonio-Sobrinho, JHEP {\bf 0411} (2004) 068; A. P. Balachandran, T. R. Govindarajan, C. Molina, P. Teotonio-Sobrinho, JHEP {\bf 0410} (2004) 072.

\bibitem{madore} J. Madore, {\it An Introduction to Noncommutative Differential Geometry and its Physical Applications}, (London Mathematical Society Lecture Note Series).

\bibitem{balbook} A.P. Balachandran, S. Kurkcuoglu and S. Vaidya, {\it Lectures on fuzzy and fuzzy SUSY physics}, World Scientific
(2007).

\bibitem{govind} T.R.Govindarajan, Pramod Padmanabhan and T.Shreecharan J. Phys.{\bf A43} (2010) 205203.
 
\bibitem{hoppe} J. Arnlind, M. Bordemann, L. Hofer, J. Hoppe and H. Shimada JHEP {\bf 0906} (2009) 047. 


\bibitem{denjoe1} Denjoe O'Connor (private communication)

\bibitem{digaltrg} S. Digal and T. R. Govindarajan, arXiv:1108.3320 [hep-th]

\bibitem{steinacker} Harold Steinacker, Nucl. Phys. {\bf B810} (2009) 1
\bibitem{ydri} Rodrigo-Delgadillo Blando, Denjoe O'Connor, B Ydri,
Phys. Rev. Letts, {\bf 100} (2008) 201601.

\bibitem{dolan} Brian P Dolan, Idrish Huet, Sean Murray, Denjoe O'Connor,
JHEP{\bf 0707} (2007) 007.

\bibitem{fuzzybh} S. Digal, T. R. Govindarajan, Kumar S. Gupta and X. Martin JHEP {\bf 1201} (2012) 027 arXiv:1109.4014 [gr-qc]. 



\end{thebibliography}
\end{document}